# High-temperature cyclic oxidation kinetics and microstructural transition mechanisms of Ti-6Al-4V composites reinforced with hybrid (TiC+TiB) networks

Shaolou Wei, Lujun Huang[*], Xinting Li, Qi An and Lin Geng

*School of Materials Science and Engineering, Harbin Institute of Technology, Harbin, 150001, P.R. China*

**Abstract**

The microstructural features and high-temperature oxidation resistance of hybrid (TiC+TiB) networks reinforced Ti-6Al-4V composites were investigated after fabricated with reaction hot pressing technique. The inhomogeneous distribution of hybrid reinforcers resulted in a sort of stress-induced grain refinement for α-Ti matrix phase, which was further facilitated by heterogeneous nucleation upon additive interfaces. HRTEM analyses revealed the crystallographic orientation relation between TiB and α-Ti phases as $(201)_{TiB}//(\bar{1}100)_{\alpha\text{-Ti}}$ plus $[11\bar{2}]//[0001]_{\alpha\text{-Ti}}$, while TiC and α-Ti phases maintained the interrelation of $(200)_{TiC}//(\bar{2}110)_{\alpha\text{-Ti}}$ and $[001]_{TiC}//[01\bar{1}0]_{\alpha\text{-Ti}}$. The hybridly reinforced Ti-6Al-4V/(TiC+TiB) composites displayed superior oxidation resistance to both the sintered matrix alloy and the two composites reinforced solely with TiC or TiB addition during the cyclic oxidation at 873, 973 and 1073 K respectively for 100 h. The hybrid reinforcers volume fraction was a more influential factor to improve oxidation resistance than the matrix alloy powder size. As temperature rose from 873 to 1073 K, the oxidation kinetics transferred from the nearly parabolic type through qusilinear tendency into the finally linear mode. This corresponded to the morphological transition of oxide scales from a continuous protective film to a partially damaged layer and ended up with the complete spallation of alternating alumina and rutile multilayers. A phenomenological model was proposed to elucidate the growth process of oxides scales. The release of thermal stress, the suppression of oxygen diffusion and the fastening of oxide adherence were found as the three major mechanisms to enhance the oxidation resistance of hybrid reinforced composites.

[*] Corresponding author: L.J. Huang (huanglujun@hit.edu.cn)

## 1. Introduction

Seeking high-performance structural materials with high strength, low density and superior oxidation resistance in the temperature range of 873-1073 K has become a worldwide challenge since the 1980s due to the rapid development in the fields of aeronautics and astronautics[1]. Titanium-based alloys are considered as the promising candidates for new-generation spacecraft applications because they possess higher specific strength when compared to high-temperature alloys and more favorable processibility together with machinability than ordered intermetallics[1, 2]. In particular, Ti-6Al-4V alloy has become the most widely applied example since its first design several decades ago[3, 4]. An abundance of studies have emerged in the literature focusing on the novel preparation and processing methods[5, 6], plastic deformation mechanisms[7] as well as microstructural optimization for this series of alloys[8-10]. In spite of the desirable overall mechanical properties, their relatively low specific stiffness, deteriorated ultimate strength and high chemical activity at elevated temperatures hinder them from further applications[11].

During the past two decades, various attempts have been made by experimentalists to enhance the mechanical properties of Ti-6Al-4V alloy so as to meet the increasing demands of aerospace vehicles. Among the numerous techniques to strengthen this alloy, introducing ceramic-based reinforcements is often regarded as the most optimal approach due to its low cost and convenience in fabrication. Dating back to the 1990s, continuous SiC fibers were employed to improve the creep resistance of Ti-6Al-4V alloy as reported by Schwenker et al[12]. To overcome the disadvantages such as the anisotropic effects of mechanical performances for this kind of composites, short TiB whiskers were subsequently used as substitutions. Plenty of investigations toward the fundamentals of preparation and processing[13, 14], mechanical properties[15, 16] as well as deformation mechanisms[17] for Ti-6Al-4V/TiB composites have been accomplished in the recent ten years. Owing to their high strength, superior hardness and stable physicochemical properties at elevated temperatures, TiC particles were also considered as another kind of promising reinforcer candidate for Ti-6Al-4V alloy[18, 19]. With the combination of both TiC and TiB, Ti-6Al-4V/(TiC+TiB) composites have been produced through various fabrication techniques, resulting in the enhancement of high-temperature mechanical properties[20-22]. The traditional Ti-6Al-4V matrix composites reinforced with TiB or TiC usually display microstructural homogeneity, whereas Huang et al[23] firstly reported that Ti-6Al-4V/TiB composites with network-distributed TiB, which were fabricated by reaction hot pressing techniques, achieved an unprecedented enhancement of high-temperature tensile properties. It was previously reviewed by Huang et al[11] that tailoring the reinforcement network distribution, which led to microstructural inhomogeneity, could also improve the mechanical properties at elevated temperature for other types of titanium-based alloys.

As one of the main modes of degradation, the oxidation of metals and alloys has attracted much attention particularly in the field of high-temperature structural materials[24-27]. The general process of oxidation-caused failure includes at least three stages: the growth of oxide scale, the compressive-stress induced crack propagation and ultimately the spallation of scales[28, 29]. During such a process, the oxidation kinetics, the phase constitutions of oxides and microstructural evolution are the main topics to be systematically explored. Up till now, a lot of research has been conducted in both experimental and theoretical aspects towards these ends for the oxidation process of titanium-based especially Ti-6Al-4V alloys[30-36]. Owing to their high chemical activity, strong affinity with oxygen and inferior oxygen diffusion resistance, the oxidation stimulated degradation becomes one of the main factors deterring titanium-based alloys from efficient applications at higher temperatures. Consequently, ceramic reinforcements with homogenous distribution were introduced to titanium alloys with another purpose to enhance oxidation resistance apart from improving mechanical properties at elevated temperatures, about which some work has been accomplished[37]. As reviewed by Huang et al[11], Ti-6Al-4V matrix composites with inhomogeneous reinforcements distribution were proved to exhibit superior high-temperature mechanical properties. Therefore, it is of great significance to examine their oxidation resistance, for which much work still remains to be done.

On the basis of previous investigations[11], network TiB reinforced Ti-6Al-4V composites show superior high-temperature tensile properties, but their oxidation resistance is destructed due to the strong affinity between TiB and oxygen. Although TiC reinforced composites with similar microstructural features possess desirable oxidation resistance, their mechanical properties are inferior to the matrix alloy, because the well-assembled TiC particles suppress the deformation compatibility. The objective of the present work is to design and fabricate hybrid (TiC+TiB) networks reinforced Ti-6Al-4V composites with various network parameters by optimizing the joint advantages that TiB and TiC respectively bring into the matrix alloy. Focuses are directed firstly to elucidate the microstructural characteristics of composites in micro and sub-micro scales and then to comparatively investigate the cyclic oxidation resistance of these composites with hybrid and single reinforcements as well as matrix alloy. The envisaged mechanisms regarding the thermodynamics and kinetics of oxide scale formation for Ti-6Al-4V/(TiC+TiB) composites are explored in details.

## 2. Experimental procedures
### 2.1 Fabrication of composite materials

Materials used in this study, including the monolithic Ti-6Al-4V alloy, Ti-6Al-4V/TiC, Ti-6Al-4V/TiB and Ti-6Al-4V/(TiC+TiB) composites, were all prepared by reaction hot pressing technique. On the grounds of previous investigations[11], spherical Ti-6Al-4V alloy powders with a diameter range from 45 to 150 μm (their chemical compositions are listed in Tab.1), fine TiB$_2$ (3 μm) and graphite (1 μm) powders were employed as raw materials for alloy and composite fabrication. Because the same processing technique was applied for both the single addition and hybrid reinforced composites, the rest of this section takes the latter as an example to illustrate the details of composites fabrication. At first, the three types of powders were mechanically ball-milled under the protection of high purity argon gas at a speed of 150 rpm for 5 h so as to adhere fine TiB$_2$ and graphite powders uniformly onto the surface of alloy powders and simultaneously avoid severe plastic deformation and oxidation. The milled mixtures were subsequently hot-pressed and sintered at 1473 K for 50 min under a pressure of 25 MPa. These were the eventually optimized parameters obtained via thermodynamic and kinetic calculations as well as experimental confirmation. They were then annealed to room temperature, during which the vacuum was kept below $10^{-3}$ Pa in order to prevent both hydrogen adsorption and oxidation. In the framework of thermodynamic design for titanium matrix composites fabrication, the following two in-situ chemical reactions took place spontaneously at temperatures beyond 1200 K, which consequently gave rise to the formation of network-distributed TiC and TiB reinforcements:

$$C(s, \text{graphite}) + Ti(s, \beta) \rightarrow TiC(s) \quad (1)$$
$$TiB_2(s) + Ti(s, \beta) \rightarrow TiB(s) \quad (2)$$

The preparation of Ti-6Al-4V alloy is quite similar to the procedures explained above, except that the alloy powders with diameters range from 85-125 μm were directly hot pressed under the same conditions without being mechanically ball-milled.

Tab. 1 Chemical compositions of Ti-6Al-4V alloy powders used for composites fabrication

| Element | Ti | Al | V | Fe | O | Si | C | N | H |
|---|---|---|---|---|---|---|---|---|---|
| Wt.% | 89.228 | 6.420 | 4.120 | 0.180 | 0.120 | 0.024 | 0.013 | 0.011 | 0.004 |

## 2.2 High-temperature cyclic oxidation

The specimens for high-temperature cyclic oxidation experiments were prepared through the following traditional techniques: cuboid-shaped samples with dimensions of 10×10×3 mm were cut from bulk φ60×25 mm alloys and composites by wire electrical discharge machining (WEDM). In order to eliminate the oxide scales and surface residual stress yielded in WEDM process, all the specimens were mechanically polished on a series of SiC abrasive paper. The specimens were then ultrasonically cleaned in ethyl alcohol bath for 5 min, and chemically etched using 10 vol.% HF solution so as to obtain the clear observation of microstructures for both matrix alloy and reinforcements. In the present work, cyclic oxidation experiments were carried out at 873, 973 and 1073K for 100h respectively. The specimens were placed inside φ15×8mm alumina crucibles and subsequently oxidized in heat treatment furnaces under laboratory air atmosphere, during which the mass change of the specimens were monitored by a PTX-FA210 high accuracy electrical balance with precision of 0.1 mg after each cycle of 10 h. Tab. 2 shows the detailed constitution parameters of the eight kinds of experimental materials prepared in this work. The first four materials were used to clarify the effects of different reinforcements on the kinetics of cyclic oxidation, while the rest were employed to systematically investigate the kinetic features and structural evolution and also to seek the possible optimization design for Ti-6Al-4V/(TiC+TiB) composites.

## 2.3 Materials characterization

The phase identifications of both as-sintered alloys and hybrid reinforced composites were performed by a Panalytical Empyrean X-ray diffractometer (XRD) with the scan angle range of 20-90 degrees in step of 0.02 degree before and after cyclic oxidation. The surface morphology and microstructural evolution of the oxidized and unoxidized specimens were investigated via Zeiss Axiovert 200 MAT optical microscope and Zeiss SUPRA 55 SAPPHIR scanning electron microscope (SEM) equipped with an INCA 300 energy dispersive spectrometer (EDS). An FEI Talos 200 transmission electron microscope (TEM) equipped with a high angle annular dark field (HAADF) detector was employed to study the interfacial characteristics and orientation relationships between reinforcements and matrix together with the elemental distributions in Ti-6Al-4V/(TiC+TiB) composites. The TEM samples were prepared by ion beam milling methods. The specimens for cross-sectional analyses were inlaid in epoxy so as to prevent oxide scales from spallation before they were processed following the conventional metallographic procedures.

Tab. 2 Material constitutions designed for cyclic oxidation experiments

| No. | Matrix size, μm | Reinforcement | Reinforcer fraction, Vol. % |
| --- | --- | --- | --- |
| 1 | 85-125 | None | None |
| 2 | 85-125 | TiC | 5 |
| 3 | 85-125 | TiB | 5 |
| 4 | 85-125 | TiC+TiB | 5 |
| 5 | 45-85 | TiC+TiB | 5 |
| 6 | 125-150 | TiC+TiB | 5 |
| 7 | 45-85 | TiC+TiB | 8 |
| 8 | 85-125 | TiC+TiB | 8 |

## 3. Results and Discussion
### 3.1 Structures of as-sintered materials

On the basis of a series of systematic studies on Ti-6Al-4V alloy[4, 36, 38], the typical morphologies of this dual-phase titanium-based alloy may be classified according to the arrangements of (α+β) phases as Widmanstätten lamellar microstructure, equiaxed microstructure and duplex microstructure. The SEM analyses of the as-sintered alloy in this work (Fig.1 (a)) reveal a typical Widmanstätten lamellar microstructure consisting of a relative large volume fraction of α-phase (dark region) and a small volume fraction of β-phase (bright region), which is the result of quasi-equilibrium annealing process.

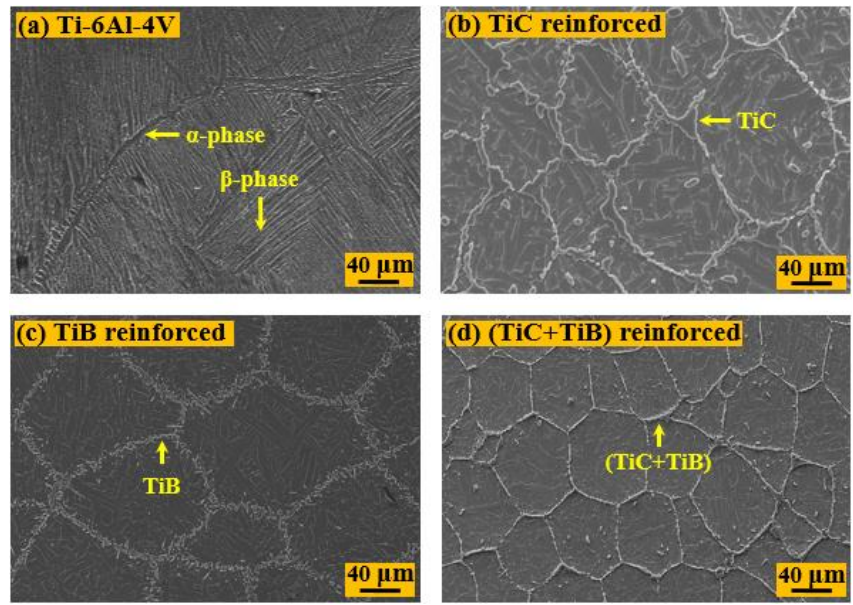

Fig.1 SEM images of as-sintered Ti-6Al-4V alloy and composites:
(a) Ti-6Al-4V; (b) Ti-6Al-4V/TiC; (c) Ti-6Al-4V/TiB; (d) Ti-6Al-4V/(TiC+TiB).

In the case of Ti-6Al-4V/TiC composites, it is clear from Fig.1 (b) that a three-dimensional network microstructure, which is made up of the closely assembled particle-shaped TiC, is successfully obtained and the network size is near the diameters of matrix alloy powders (85-125 μm). As can be seen from Fig.1 (c), TiB whiskers distribute selectively on the boundaries of Ti-6Al-4V powders, which also form network microstructures. Similarly, Ti-6Al-4V/(TiC+TiB) composites exhibit network-distributed reinforcements with equal volume fractions of TiC particles and TiB whiskers. Once comparing the microstructure of monolithic Ti-6Al-4V alloy with that of the alloy matrix within composites, it is found that the microstructure transfers from Widmanstätten lamellar microstructure into refined equiaxed microstructure, which has proved to be beneficial for the mechanical performances especially at elevated temperatures[11]. It is worthy of being noticed that a relatively small number of TiC particles also distribute inside the alloy matrix, as displayed in Fig. 1 (b) and (d). Since graphite was used in the present work to induce the formation of in-situ TiC, it is possible that the carbon atoms possess so high mobility at 1473 K sintering temperature that they partially diffuse into the inner part of alloy matrix, which consequently results in the localized TiC formation.

The XRD patterns of typical Ti-6Al-4V/(TiC+TiB) composites together with the as-sintered Ti-6Al-4V alloy are both provided in Fig. 2. It is evident that TiC and TiB reinforcers are indeed formed inside the composites, whereas there are no diffraction peaks of graphite and $TiB_2$, which are the reactants of Equ.s (1) and (2). This indicates that these two in-situ chemical reactions took place completely under the optimized parameters for reaction hot pressing. To combine the structural merits of TiC particles and TiB whiskers, the ideal network microstructure designed for Ti-6Al-4V/(TiC+TiB) composites is illustrated in Fig. 2(b). The SEM micrographs in Fig.s 2(c) and (d) confirm the successful accomplishment of such a structural design. TiC particles assemble with each other and act as the "grain boundaries" separating Ti-6Al-4V alloy matrix nearby, while the needle-shaped TiB whiskers insert into the alloy matrix at each side of boundaries as dowels.

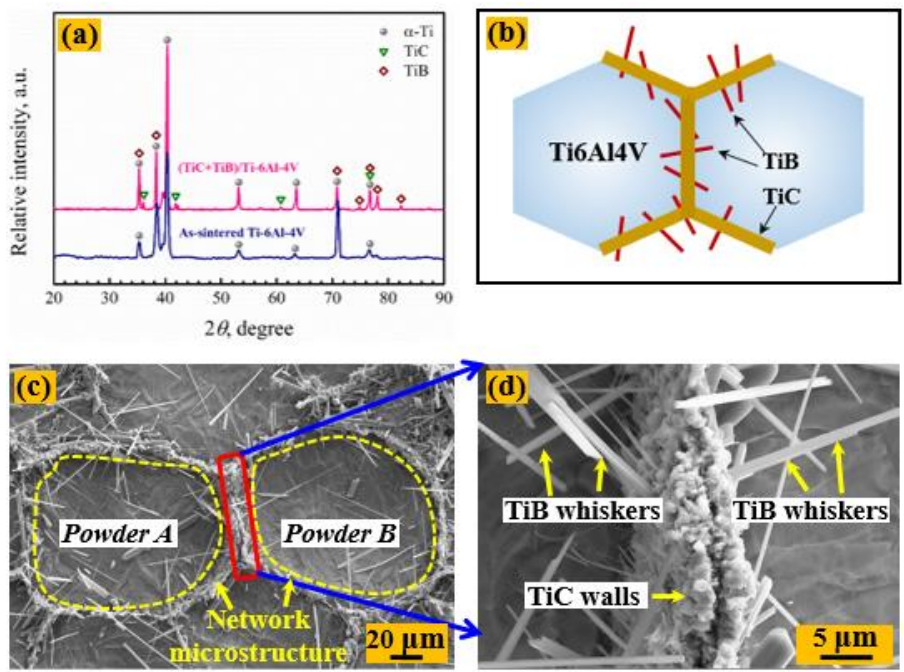

Fig.2 Phase identification and microstructural characterization of Ti-6Al-4V/(TiC+TiB) composites:
(a) XRD patterns; (b) schematic of microstructural design;
(c) microstructure after deep etching; (d) magnified network microstructure.

As previously addressed, the introduction of TiB and TiC reinforcements induces the transition of Widmanstätten lamellae into refined equiaxed microstructure. Such effects are more evidently observed in the OM micrograph of Ti-6Al-4V/(TiC+TiB) composites presented in Fig. 3(a). To further illustrate the possible mechanisms of α-phase refinement, a phenomenological model is proposed in Fig. 3(c). Suppose that in the cooling process of reaction hot pressing, three original α grains A, B and C in the matrix alloy possess the tendency of growing in a direction marked with the bold arrows until they come in contact with the reinforcements. Since the modulus of Ti-6Al-4V alloy is about 300 GPa lower than those of TiC or TiB, the network structured reinforcements act as "rigid walls" which suppress the growth of α grains by exerting compressive stress $\sigma_g$ on those grains. This stress may be well preserved to room temperature and consequently becomes residual stress between reinforcements and matrix alloy, which results in the stress-expedited chemical corrosion of matrix alloy as seen in Fig. 3(b). This model could also well explain the reason why α grains located closer to the reinforcements region have smaller sizes than those in the central part of alloy matrix. Except for the stress between reinforcements and matrix alloy, another factor might also bring about the same consequence. During the β→α solid phase transformation, the TiB and TiC are possible to become the heterogeneous nucleation sites for α phase. Therefore, the nucleation rate at their interfaces is more than one order of magnitude larger than homogenous nucleation, which results in refined equiaxed α-phase.

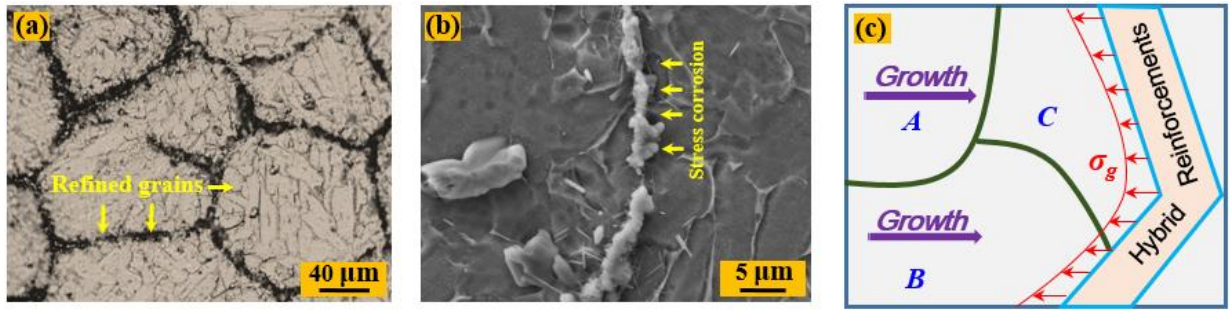

Fig. 3 Reinforcements induced grain refinement of matrix Ti-6Al-4V alloy:
(a) OM micrograph of refined grains; (b) stress-expedited chemical corrosion along network boundaries;
(c) phenomenological model for grain refinement.

The TEM studies shown in Fig. 4 reveal that both TiB and TiC reinforcers bond well with the matrix α-Ti phase, and the interfaces between ceramic reinforcements and matrix alloy are clear and free from extra interfacial reaction products or precipitations. Here the selected area diffraction patterns of TiB and TiC are presented as inserted figures. Moreover, TiB and matrix α-Ti phase display an orientation relationship extremely close to $(201)_{TiB}//(\bar{1}100)_{\alpha\text{-Ti}}$; $[11\bar{2}]//[0001]_{\alpha\text{-Ti}}$, while TiC and α-Ti phase most probably hold an orientation relationship of $(\bar{2}00)_{TiC}//(\bar{2}110)_{\alpha\text{-Ti}}$; $[001]_{TiC}//[01\bar{1}0]_{\alpha\text{-Ti}}$. This is obviously demonstrated by the fast Fourier transformation analyses of the HRTEM images as demonstrated in Fig. s 4 (c) and (f).

Fig. 5 presents the HAADF images corresponding to the elemental distribution analyses of reinforcements in the coupled growth region. The alloying element Al exhibits a remarkable depletion zone in the region where TiC or TiB exists, while Ti and V display fairly uniform distributions. When comparing Fig. 5 (c) with (g), it is found that B becomes enriched in a relatively small area, which is probably the evidence for the presence of TiB whiskers. This strongly supports the assumption that heterogeneous nucleation took place during β→α phase transformation, which is previously considered as one of the possible mechanisms for α-phase refinement.

### 3.2 Cyclic oxidation process

Due to the complexity of phase constitutions within Ti-6Al-4V/(TiC+TiB) composites, it is necessary to predict the possible oxidation reactions and products via theoretical calculations in advance. Since the cyclic oxidation experiments in the present work are conducted at constant temperature and pressure, according to the principles of classical thermodynamics, the variation of Gibbs free energy is the most efficient criterion to describe the possibility whether or not a chemical reaction can happen. Deriving from the first law of thermodynamics and the general correlation between state functions as well as the series for heat capacity at constant pressure, the standard molar Gibbs free energy change is expressed as the Gibbs-Helmholtz formula[39]:

$$\Delta G_r = A + BT - 4.1868\left(\Delta aT \ln T + 5\times 10^{-4} \Delta bT^2 + 5\times 10^4 \Delta cT^{-1}\right) \qquad (3)$$

where $\Delta G_r$ represents the standard molar Gibbs free energy change for a specific chemical reaction, $\Delta\delta$, $\Delta a$, $\Delta b$ and $\Delta c$ are coefficients derived from the series of heat capacity, $A$ and $B$ are both integration constants which are determined using the data of standard molar formation Gibbs energy $\Delta G_f$ and enthalpy change $\Delta H_f$ at 298 K.

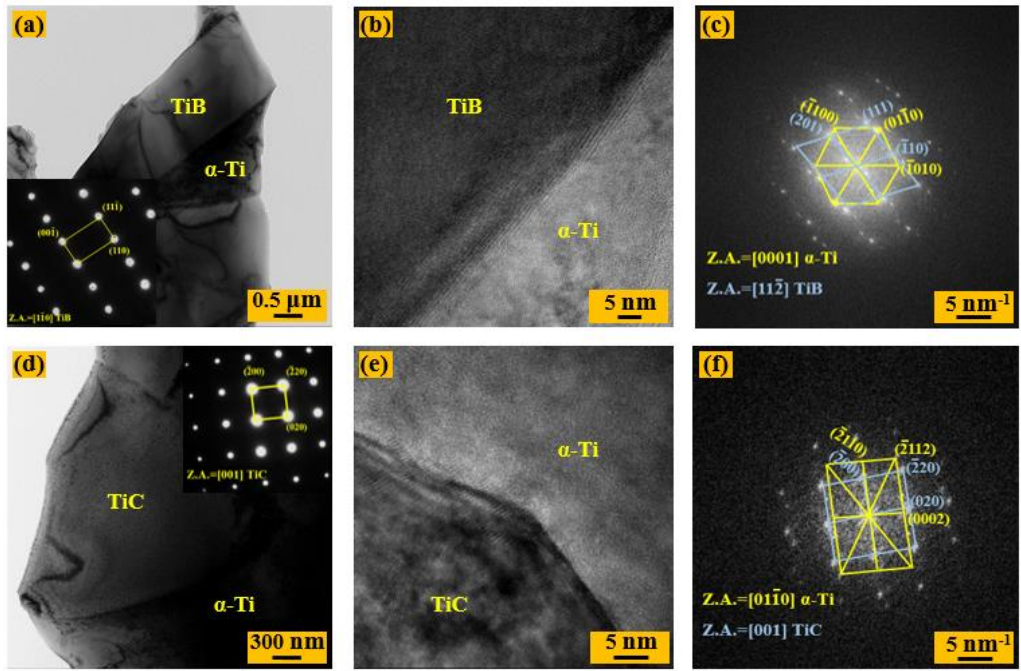

Fig. 4 TEM images of interfaces within Ti-6Al-4V/(TiC+TiB) composites:
(a) BF image of TiB/α-Ti region; (b) HRTEM image of TiB/α-Ti interface; (c) FFT of (b);
(d) BF image of TiC/α-Ti region; (e) HRTEM image of TiC/α-Ti interface; (f) FFT of (e).

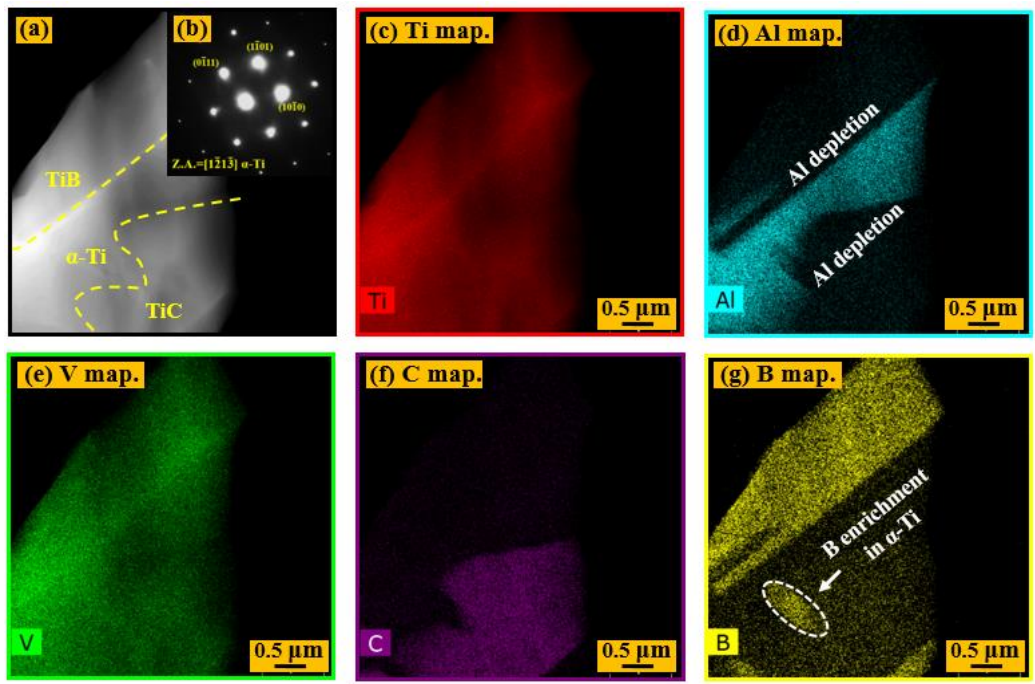

Fig. 5 STEM-EDS analyses of TiB/α-Ti/TiC regions:
(a) and (b) HAADF and BF region of selected areas;
(c)-(g) EDS elemental mapping of Ti, Al, V, C and B distributions respectively.

Because the oxidation time of Ti-6Al-4V alloy and composites is relatively long (100 h in total and 10 h for each cycle), it is reasonable to assume that the main resultants are the high valence oxides which also have relatively stable physicochemical properties. All the possible chemical reactions in the temperature range of 873-1073 K are listed as follows, and the thermodynamic parameters used in the calculations are summarized in Tab.3.

$$\frac{1}{2}Ti(s) + O_2(g) \rightarrow \frac{1}{2}TiO_2(s, anatase) \quad (4)$$

$$\frac{1}{2}Ti(s) + O_2(g) \rightarrow \frac{1}{2}TiO_2(s, rutile) \quad (5)$$

$$\frac{4}{3}Al(s) + O_2(g) \rightarrow \frac{2}{3}Al_2O_3(s) \quad (6)$$

$$\frac{4}{5}V(s) + O_2(g) \rightarrow \frac{2}{5}V_2O_5(s) \quad (7)$$

$$\frac{1}{2}TiC(s) + O_2(g) \rightarrow \frac{1}{2}TiO_2(s) + \frac{1}{2}CO_2(g) \quad (8)$$

$$\frac{4}{7}TiB(s) + O_2(g) \rightarrow \frac{4}{7}TiO_2(l) + \frac{2}{7}B_2O_3(l) \quad (9)$$

The calculation results of Eqs. (4-9) are demonstrated in Fig. 6 (a) and (b). The $\Delta G_r$ values of all the reactions listed above increase almost linearly within the temperature range of 873-1073 K and also show negative values, which suggests that these reactions should be thermodynamically possible to take place during cyclic oxidation. According to Fig. 6(a), among all the alloying elements, Al exhibits the strongest thermodynamic tendency of being oxidized, while V shows the least oxidation probability since the absolute value of $\Delta G_r$ for Eq. (6) is much larger than that of Eq. (7). Note that such results predict that Ti-6Al-4V alloy is likely to be selectively oxidized in the temperature range of 873-1073 K. This is in agreement with the investigation about the isothermal oxidation behavior of Ti-6Al-4V reported by Guleryuz et al[35]. Fig. 6 (b) reveals that in terms of classical thermodynamics, TiC displays more desirable oxidation resistance than TiB as well as the alloying elements Ti and Al, which contributes to one of the reasons why TiC is considered as a promising reinforcement candidate to enhance high-temperature performances for Ti-base alloys[40, 41].

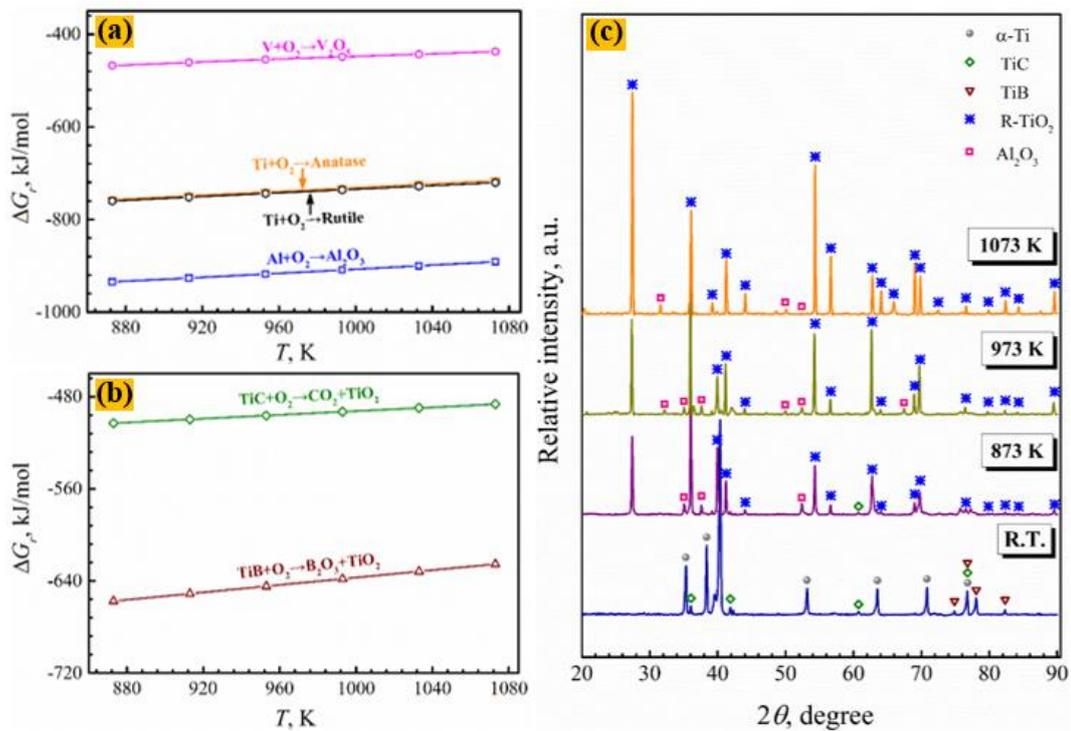

Fig. 6 Oxidation products of Ti-6Al-4V/(TiC+TiB) composites at different temperatures for 100 h:

(a) and (b) calculated $\Delta G_r$ for possible oxidation reactions;

(c) XRD patterns before and after oxidation.

Tab. 3 Thermodynamic parameters used for physicochemical calculations[39]

| Component | $a$, J/mol·K | $b$, J/mol·K$^2$ | $c$, J·K/mol | $\Delta H_f$, kJ/mol | $\Delta G_f$, kJ/mol |
|---|---|---|---|---|---|
| α-Ti | 5.28 | 2.40 | 0.00 | 0.00 | 0.00 |
| Al | 4.94 | 2.96 | 0.00 | 0.00 | 0.00 |
| V | 4.90 | 2.58 | 0.20 | 0.00 | 0.00 |
| TiC | 11.93 | 0.02 | 0.00 | -184.10 | -176.88 |
| TiB | 12.91 | 0.00 | 0.00 | -160.25 | -149.90 |
| O$_2$ | 7.16 | 0.00 | 0.00 | 0.00 | 0.00 |
| TiO$_2$ (A) | 17.83 | 0.50 | -4.23 | -944.75 | -929.75 |
| TiO$_2$ (R) | 17.79 | 0.28 | 0.00 | -933.03 | -918.15 |
| Al$_2$O$_3$ | 25.48 | 4.25 | -6.52 | -1678.20 | -1540.00 |
| B$_2$O$_3$ | 30.45 | 0.00 | 0.00 | -1272.50 | -1192.90 |
| V$_2$O$_5$ | 46.54 | -3.90 | -13.22 | -1551.3 | -1454.12 |

Fig. 7 presents the experimentally measured cyclic oxidation kinetics curves at 873, 973 and 1073 K. It is clear that hybrid (TiC+TiB) reinforced composites possess superior oxidation resistance to single TiC or TiB reinforced composites and Ti-6Al-4V alloy. At the relatively low temperature of 873 K in Fig. 7 (a), the mass-gain kinetics for Ti-6Al-4V alloy and three kinds of composites follow quasi-parabolic principles, indicating that protective oxide scales were formed and free from macro-scale spallation during cyclic oxidation, which has been confirmed by the macroscopic observations. Statistical data show that after cyclically oxidized for 100 h, the total mass-gain of hybrid reinforced composites is 8.1 %, 8.0 % and 4.0 % lower than those of Ti-6Al-4V alloy, Ti-6Al-4V/TiB and Ti-6Al-4V/TiC composites, respectively.

As experimental temperature rises up to 973 K, the mass-gain kinetics of Ti-6Al-4V transfers into a quasi-linear trend, while the masses of three types of composites increase linearly as oxidation time extends from 20 to 70 h and subsequently follow an inconspicuous parabolic law in the last 3 cycles. The macroscopic observation of surface morphology reveals that the oxide scale on Ti-6Al-4V alloy displays the most severe spallation, while the oxide scale over hybrid reinforced composites adheres well onto the alloy substrate. Additionally, the final stage mass-gain values at 973 K are generally one order of magnitude higher than those of 873 K. Among all the oxidized specimens, hybrid reinforced composites still show the least mass-gain, which is 33.9 %, 18.9 % and 10.5 % lower than Ti-6Al-4V alloy, Ti-6Al-4V/TiB and Ti-6Al-4V/TiC composites, respectively.

Once cyclic oxidation temperature reaches 1073 K, the mass-gain kinetic curves of all the materials exhibit conspicuous linear trend, as can be seen in Fig. 7(c). In such a process, the oxide scales over Ti-6Al-4V alloy and the three types of composites have desquamated after each cycle, where Ti-6Al-4V/(TiC+TiB) composites possess the least amount of spallation. Similar to the situations of 873 and 973 K, the mass-gain of hybrid reinforced composites is 36.3 %, 12.3 % and 16.4 % lower as compared with those of Ti-6Al-4V alloy, Ti-6Al-4V/TiB and Ti-6Al-4V/TiC composites. It is adequately inferred that the introduction of hybrid (TiC+TiB) reinforcers into Ti-6Al-4V alloy results in the enhanced oxidation resistance as compared with Ti-6Al-4V alloy and those two composites reinforced solely with TiB or TiC, and this effect becomes more prominent with the rise of cyclic oxidation temperature. A comprehensive insight into the macroscopic observation of surface morphological evolution and the mass-gain trend indicates that a possible correlation might exist between the spallation of oxide scale and the cyclic oxidation kinetics.

In order to further clarify the effects of reinforcements volume fraction and alloy matrix size on the oxidation behavior and to explore whether or not microstructural optimization can be realized, Ti-6Al-4V/(TiC+TiB) composites with various constitution parameters were cyclically oxidized at the same temperatures for 100 h, which is demonstrated in Fig.8. The variations of overall kinetic features in the experimental temperature range is rather similar to the characteristics illustrated in Fig. 7.

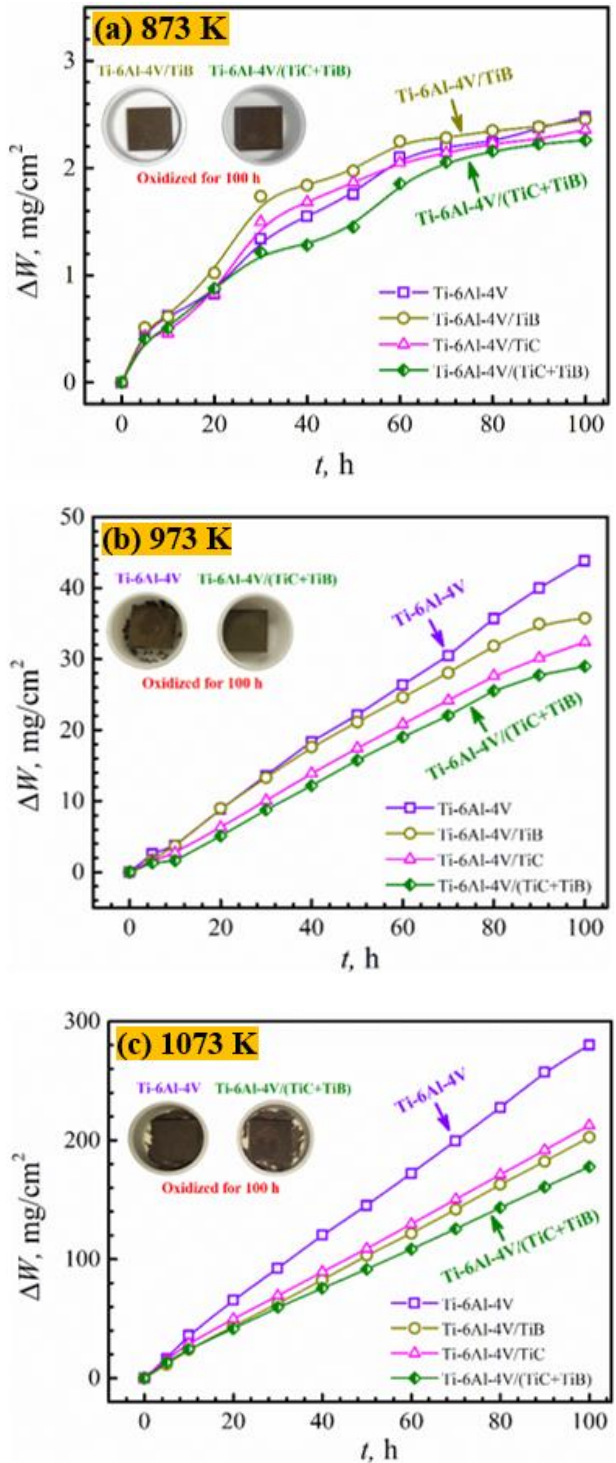

Fig. 7 Cyclic oxidation kinetics of Ti-6Al-4V alloy and composites at various temperatures:
(a) 873 K; (b) 973 K; (c) 1073 K.

A kinetic transition proceeds from the quasi-parabolic type at 873 K to the quasi-linear combined with atypical parabolic type at 973K and eventually to the conspicuous linear mode at 1073 K. Among all the hybrid reinforced composites, the one with matrix size ranging from 85-125 μm and 5 Vol.% reinforcement fraction exhibits the poorest cyclic oxidation resistance at all the three experimental temperatures, where the spallation of oxide scale is also the most severe. It can also be seen from Fig. 8 that, for those composites with the constant 5Vol. % reinforcement fraction, the one with the smallest 45-85 μm matrix size possesses superior cyclic oxidation resistance, which is more prominent at the higher temperatures of 973 and 1073 K. Moreover, the kinetic features of the two composites with larger matrix size becomes much more alike as temperature rises. When the reinforcements volume fraction increases from 5 % to 8 %, the effect mentioned above still holds for 1073 K and 873 K, but it reverses when oxidized at 973 K. If comparing the kinetics curves for those composites with the same matrix size of 85-125 μm, it is found that the increment in reinforcement volume fraction brings in an evident enhancement in cyclic oxidation resistance, especially at higher temperatures. When the alloy matrix size decreases to 45-85 μm, even though such effects show no longer force for 873 and 973 K, the tendency in which the kinetic features evolve still demonstrates that those composites with higher reinforcement volume fraction are promising to achieve superior cyclic oxidation resistance. The reason why such a reversion exists may be ascribed to the variation in interface properties induced simultaneously by both matrix size and reinforcements volume fraction, which needs to be further explored.

Fig. 6(c) provides the XRD patterns for typical hybrid reinforced composites before and after 100 h cyclic oxidation. As previously predicted by the theory of classical thermodynamics, the oxides are mainly rutile-$TiO_2$ and $Al_2O_3$, which partially agrees with the results in the literature[35, 37]. Different from the phase constitutions of isothermally oxidized Ti-6Al-4V alloy (at 873 K and 723 K for 48 h and 12 h) reported by Guleryuz et al[35], the diffraction peaks of anatase-$TiO_2$ do not appear in the present work. This is because during a long period of 100 h cyclic oxidation at higher temperatures, the metastable anatase-$TiO_2$ could completely transform into stable rutile-$TiO_2$ phase. The absence of the $V_2O_5$ oxide of alloying element V from XRD patterns is attributed to the following two possible reasons. On the one hand, according to the thermodynamic calculations, the strong selective oxidation of Al or Ti element is likely to take place at elevated temperatures, which as a result protects V from being oxidized.

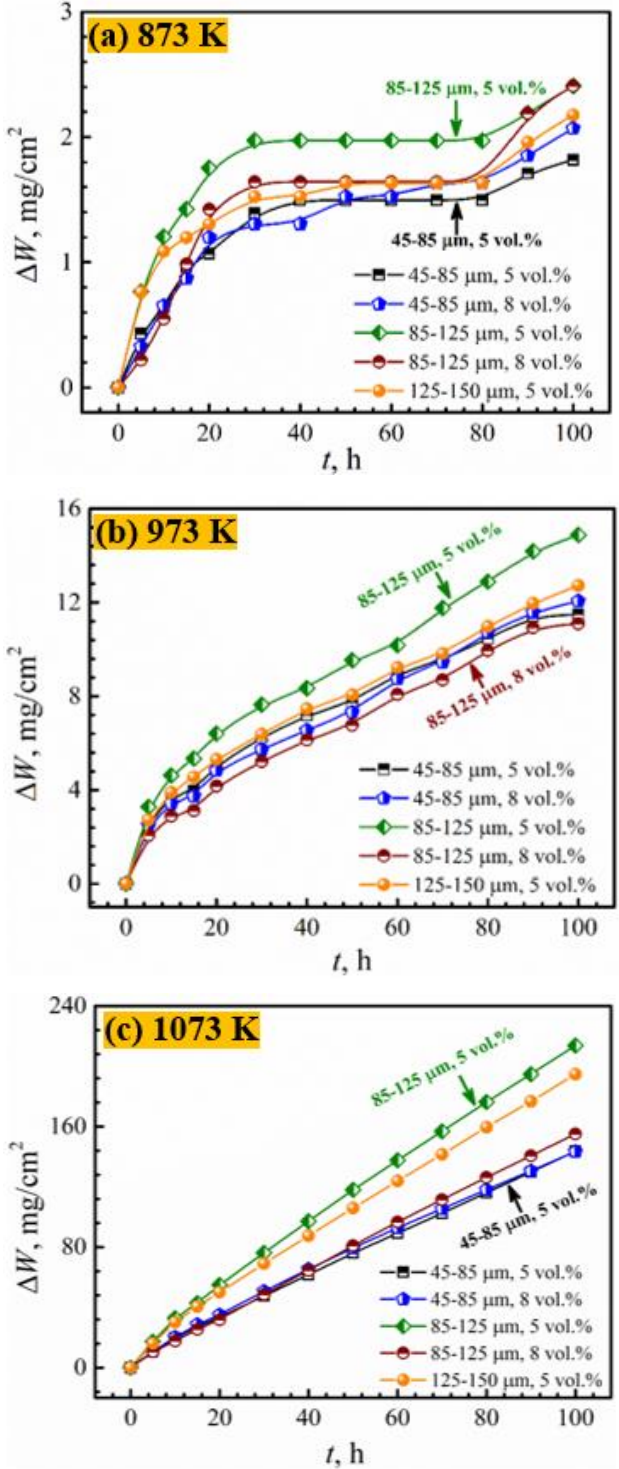

Fig. 8 Effects of materials constitutions on cyclic oxidation kinetics of Ti-6Al-4V/(TiC+TiB) composites versus temperatures: (a) 873 K; (b) 973 K; (c) 1073 K.

On the other hand, even if some amount of V is oxidized, the content of $V_2O_5$ oxide is too small to be detected by XRD methods. As can be seen from the XRD patterns of composites oxidized at 873 K for 100 h, the diffraction patterns of TiC still exist, whereas the isolated peaks of TiB vanish from the patterns. Such a phenomenon may be well explained through the thermodynamic calculation results shown in Fig. 6 (b). It is worth noting that the diffraction peaks of $B_2O_3$, which is the oxidation products of TiB, also disappear from all the patterns. Currently there are three reasonable mechanisms for this phenomenon reported in the literature: the evaporation of $B_2O_3(l)$[42]; successive reaction between $B_2O_3(l)$ and $H_2O$ (g) leading to the formation of $H_3BO_3(g)$[43]; amorphous $B_2O_3(s)$ in cooling process[44]. According to a recent study on the evaporating products of $B_2O_3$ (l) reported by Sasaki et al[45], the $B_2O_3$ (l) might not be possible to evaporate under the experimental conditions of the present work. The traces of pores formed in the composites are demonstrated in the SEM image of the oxidized composites (Fig. 9(c)), indicating that the evaporation of oxidation products does exist in the composites. Since the oxidation products of TiC also include volatile gas $CO_2$, while amorphous $B_2O_3$ powders are too small to be observed, the latter two mechanisms are both possible to result in the absence of $B_2O_3$ peaks from XRD patterns.

### 3.3 Morphological evolution of oxidized structures

The SEM micrographs of the surface morphologies for typical Ti-6Al-4V/(TiC+TiB) composites (45-85 μm, 5Vol.%) cyclically oxidized at 873 K are presented in Fig. 9. After short period of 10 h oxidation, it is found from Fig. 9 (a) that the network microstructure can still be clearly observed, even though the typical features of TiC particles and TiB whiskers vanished due to the oxidation reactions as described in Equ. s (8) and (9). The traits of β-phase of matrix Ti-6Al-4V alloy exhibits more evidently that α-phase has inferior oxidation resistance, which agrees with the results in the literature[38]. When the cyclic oxidation time extends to 100 h, the boundaries of the network (four regions marked as A, B, C and D in Fig. 9(b)) still preserve their geometrical convex features, which proves indirectly that the oxide scale over the composites is divided into smaller units. Such a microstructure might release the thermal stress and growth stress during cyclic oxidation. Fig. 9 (c) demonstrates the formation of pores within the composites, where the most probable formation mechanism is the evaporation of $H_3BO_3$ (g) as previously explained.

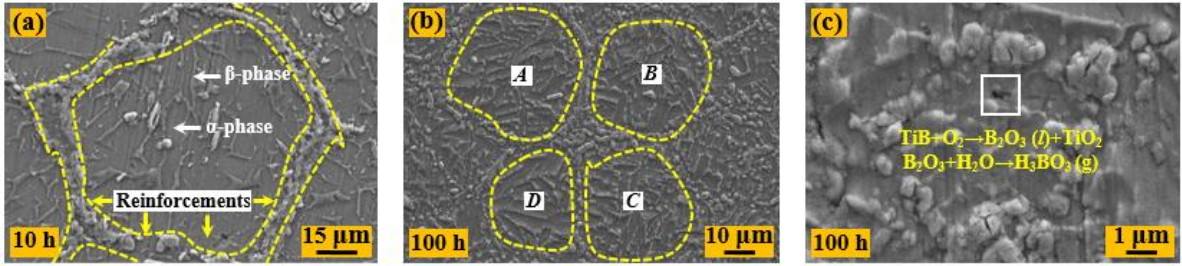

Fig .9 Surface SEM images of Ti-6Al-4V/(TiC+TiB) composites oxidized at 873 K:
(a) and (b) reinforcement networks after oxidation for 10 h and 100 h; (c) pores formed within composites.

As the temperature rises up to 973 K, the outer spallation-free region consists of oxides with two geometric morphologies: quasi-equiaxed particles and short pine-shaped needles (Fig. 10(a)). Judging from the EDS mapping results and the XRD patterns displayed in Fig. 6 (c) together with the relevant investigations in the literature[1, 31], it is concluded that $Al_2O_3$ (particle-shaped) and rutile-$TiO_2$ (needle-shaped) are concluded to form the oxide scales. If comparing the elemental distribution within the spallation-free region and the spalled zone, the alloying element Al enriches at the outermost layer (where Ti content is relatively low), while Ti enriches in the spallation area and the crack edge (marked with white arrows), which coincides with the previous studies[37, 46].

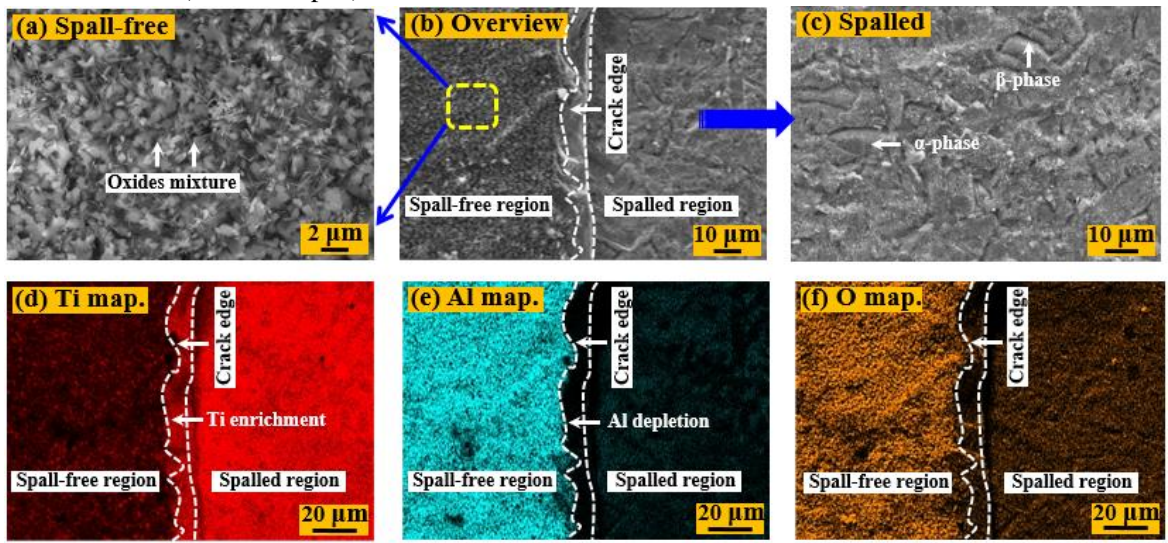

Fig .10 Surface SEM-EDS analyses of Ti-6Al-4V/(TiC+TiB) composites oxidized at 973 K for 100 h:
(a) oxide scale spallation-free region;(b) overview of typical surface morphology;
(c) oxide scale spallation region; (d)-(f) EDS elemental mapping of Ti, Al and O distributions.

Fig. 11(a) shows that mandarin-shaped rutile-$TiO_2$ was formed at the outermost spallation free region, while the inner substrate also shows traits of being oxidized. It is worth noting from Fig. 11(c) that the convex reinforcements boundaries of network A, B and C could be clearly observed. Since they locate in the spallation region, such features possibly become a direct evidence that the oxide scale is segmented into cells whose sizes are close to those of the networks. Furthermore, the network-distributed reinforcements act as dowels to fasten the oxide scale tightly onto the substrate, which suppresses the spallation of oxide scale and leads to the lower mass-gain values during cyclic oxidation. This is confirmed by the relevant macroscopic images and kinetic curves.

The cross-sectional SEM images together with EDS elemental distribution analyses for Ti-6Al-4V/(TiC+TiB)

composites after cyclically oxidized at 873, 973 and 1073 K are displayed in Fig. 12. At the lower 873 K temperature, the oxide scale with the thickness less than 5 μm adheres tightly to the alloy substrate. It acts as a reaction barrier protecting the inner part of the specimen from being oxidized, which possibly results in the quasi-parabolic mass-gain kinetics as presented in Fig.s 7 (a) and 8 (a). By comparing the elemental distribution characteristics shown in Fig.s 12 (b) and (c), it is found that the alloying element Al localizes at the outmost part of the oxide scale, indicating the existence of $Al_2O_3$, while it dilutes in the sub-layer as marked with arrows in Fig. 12(c). Referring to the results provided in Fig.s 5 (a), (f) and (g), the Al depletion area in the substrate coincides with the traits of partially or even un-oxidized reinforcements.

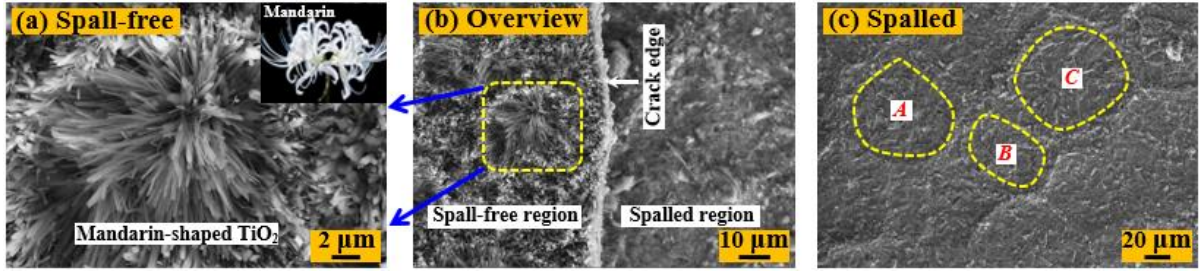

Fig. 11 Surface SEM micrographs of Ti-6Al-4V/(TiC+TiB) composites oxidized at 1073 K for 100 h:
(a) oxide scale spallation-free region; (b) overview of typical surface morphology; (c) oxide scale spallation region.

The thickness of the oxide scale increases evidently as cyclic oxidation temperature rises up to 973 K, which is shown in Fig. 12 (b). Similarly, $Al_2O_3$ locates at the outermost part of the oxide scale, resulting in the depletion of Al in nearby region. Different from the situation at 873 K, it is also found in Fig. 12 (b2) that multi-layered microstructure forms over the composites, which is the characteristics for crack formation or oxide scale spallation. This results in the deterioration of protection for oxide scale, which is likely to cause the kinetics transition in mass-gain kinetics as previously discussed. Such a phenomenon is more prominent when temperature reaches 1073 K. If comparing Fig. 12 (c) with the experimentally measured cyclic oxidation mass-gain curves presented in Fig. 8 (c), it is obvious that the number of layers within the oxide scale is exactly the same as the number of the cycles with a time period of 10 h. This consequently shifts the mass-gain kinetics into conspicuous linear type. Of all the oxidation features discussed above, it is now clear that a correlation between oxide scale spallation and kinetics transition does exist during cyclic oxidation.

### 3.4 Kinetic mechanisms of cyclic oxidation

The experimental results demonstrated above indicate that the spallation of oxide scale is a dominant factor, which affects both the cyclic oxidation kinetics and the microstructure evolution. The possible factors which induce the failure of oxide scale include the thermal stress, orientation mismatch stress, chemical composition variation, recrystallization stress and growth stress etc. Here the thermal stress is often considered as the most significant effect during cyclic oxidation. Even if the oxidized specimens are cooled at equilibrium state, it is still caused by the discrepancy in thermal expansion coefficients between oxide scale and inner substrate, let alone the conspicuously changing temperature during cyclic oxidation experiments. Because the thermal stress in such a process is often difficult to monitor by conventional experimental methods, the present work employs a dual thin-scale model proposed by Timoshenko[47] to empirically predict the thermal stress during cyclic oxidation. The expression of thermal stress between the oxide scale and substrate is derived from the principles of elastic mechanics as[47]:

$$\sigma_{Ox} = \frac{-(\alpha_{Ox} - \alpha_M)\Delta T}{\frac{2\delta_{Ox}(1-\mu_M)}{\delta_M E_M} + \frac{(1-\mu_{Ox})}{E_{Ox}}} \quad (10)$$

where $\alpha_{Ox}$, $\alpha_M$, $\delta_{Ox}$, $\delta_M$, $\mu_{Ox}$, $\mu_M$, $E_{Ox}$, $E_M$ and $\Delta T$ represent the thermal expansion coefficient, thickness, Poisson ratio, Young's modulus of the oxide scale and substrate and the temperature difference between oxidation and room temperature, respectively. Since the experimental data of thermal expansion coefficients for the composites are currently still in lack, the present work applies Kerner's model[47] to provide an empirical estimation, the form of which is expressed in Eq. (11):

$$\alpha_c = \sum_i^n K_i \cdot V_i \cdot \alpha_i \Big/ \sum_i^n K_i \cdot V_i \quad (11)$$

where $K_i$, $V_i$ and $\alpha_i$ are the bulk modulus, volume fraction and thermal expansion coefficient for $ith$ phase. According to the microstructural characteristics of oxide scale (Fig. 12), it is self-consistent to assume that the failure of oxide scale mostly initiates at the interface between sub-layered rutile-$TiO_2$ and inner substrate. The failure criterion could therefore be expressed as a function of $h/a$, which is the ratio of the oxide scale thickness to the curvature radius (Eq. (12)). The parameters used in the calculations are summarized in Tab. 4.

$$\sigma_c = 1.22 \frac{E_{Ox}}{1-\mu_{Ox}^2}\left(\frac{h}{a}\right)^2 \quad (12)$$

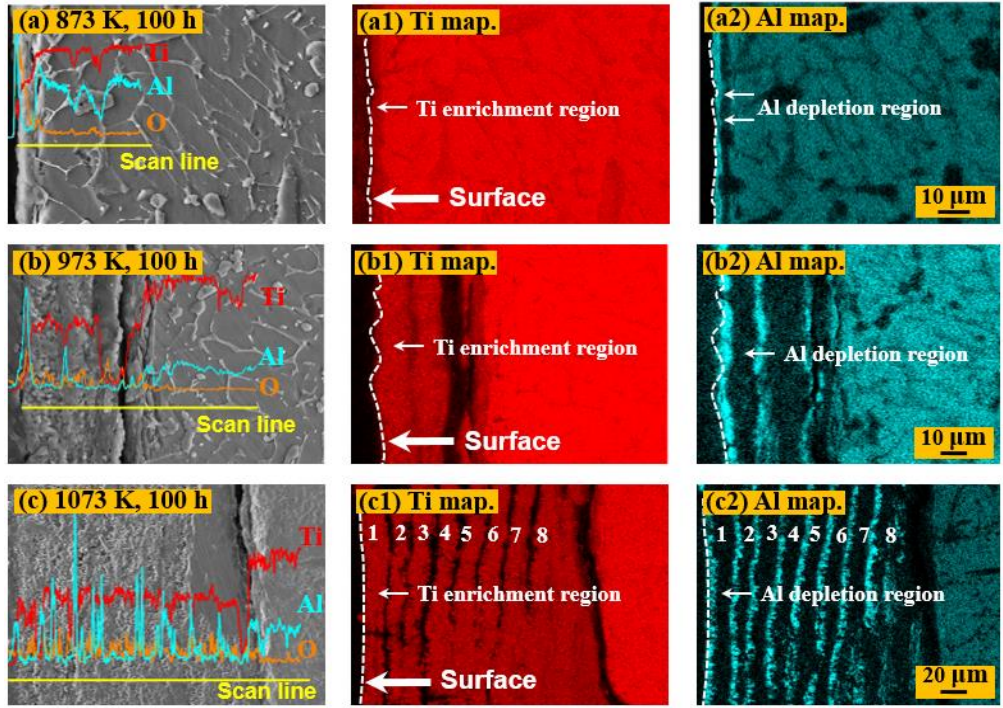

Fig. 12 Cross-sectional SEM-EDS analyses of Ti-6Al-4V/(TiC+TiB) composites oxidized at different temperatures for 100h

(a)-(c) SEM images with O and Al scans in oxide scale of composites oxidized at 873, 973 and 1073 K;

(a1)-(c1) EDS elemental mapping of Ti distribution; (a2)-(c2) EDS elemental mapping of Al distribution.

As illustrated by the calculated results in Fig. 13, it is apparent that the thermal stress within the monolithic Ti-6Al-4V alloy, Ti-6Al-4V/TiB, Ti-6Al-4V/TiC and Ti-6Al-4V/(TiC+TiB) composites increases linearly with elevated temperature during cyclic oxidation. At 873 K, the thermal stress in all materials is lower than the failure critical value $\sigma_c$, indicating that the probability to initiate cracks or spallation is relatively small. This therefore leads the protective oxide scales to be well preserved during cyclic oxidation. Such results support the quasi-parabolic mass-gain kinetics which has been previously discussed in Fig. 7(a). When temperature rises up to 973 K, the calculation results show that the thermal stress within Ti-6Al-4V alloy increases beyond the critical value, whereas the thermal stress in composites is quite close to the failure stress, indicating that cracks or spallation are more likely to form in the monolithic Ti-6Al-4V alloy than the composites. That is to say, the oxide scale over Ti-6Al-4V alloy completely loses its protection because thermal stress brings about mechanical spallation. Meanwhile, the oxide scales on composites maintain their ability to suppress successive oxidation to some extent, even though other factors such as the growth stress and orientation mismatch may contribute to the total stress value which consequently facilitates the spallation process. This could be the main reason why the mass-gain kinetics of Ti-6Al-4V alloy yields a linear trend, whereas that of composites exhibits the combined features of both inconspicuous parabolic and linear modes. Since the thermal stress in all four materials prominently exceeds the critical point, the oxide scale experiences macro-scale spallation which results in the linear mass-gain kinetics, as confirmed in Fig. 7(c). The effects of reinforcements volume fraction on the thermal stress is presented in Fig.7 (c), manifesting that Ti-6Al-4V/(TiC+TiB) composites with higher reinforcements content possess enhanced cyclic oxidation resistance, which agrees with the experimental results demonstrated in Fig. 8. The reason why the mass-gain kinetics of Ti-6Al-4V/TiC and Ti-6Al-4V/(TiC+TiB) composites contradicts with the thermal stress calculation results may be attributed to their microstructure differences. In addition to hives-shaped TiC, Ti-6Al-4V/(TiC+TiB) composites also contain whiskers-shaped TiB, which are likely to act as dowels inserting into the oxide scale, thus providing extra contribution to the suppression of oxide scale spallation. Combining the experimental results and theoretical calculations it could now be concluded that the mechanisms for the enhanced cyclic oxidation resistance of Ti-6Al-4V/(TiC+TiB) composites involve the following four aspects: (1) the introduction of TiC and TiB reinforcements into Ti-6Al-4V alloy matrix decreases the thermal stress between oxide scale and alloy substrate; (2) the network-distributed reinforcements divide the oxide scale into smaller units; (3) the closely assembled TiC particles serve as diffusion barrier for oxygen; and (4) TiB whiskers insert into the oxide scale providing extra fastening stress.

Tab. 4 Physical parameters used to calculate thermal stress[39]

| Materials | $\alpha$, $10^{-6}$ K$^{-1}$ | $K$, GPa | $E$, GPa | $\mu$ |
|---|---|---|---|---|
| Ti-6Al-4V | 11.36 | 113.13 | 112.00 | 0.34 |
| TiB | 8.60 | 205.50 | 452.50 | 0.15 |
| TiC | 7.95 | 250.00 | 446.00 | 0.20 |
| R-TiO$_2$ | 7.85 | 166.70 | 230.00 | 0.27 |

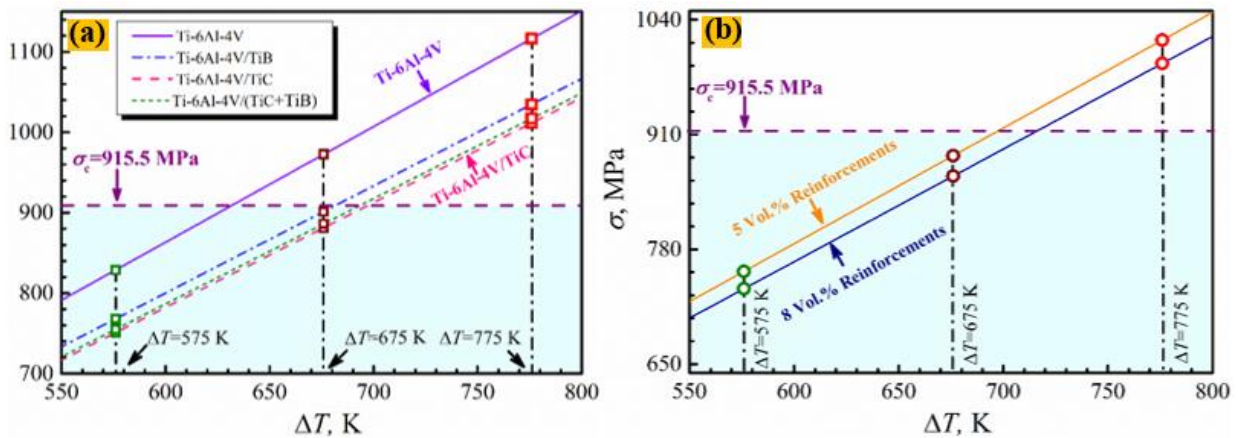

Fig. 13 Calculation results of thermal stress during cyclic oxidation:

(a) thermal stress versus temperature difference for four types of materials;

(b) effect of reinforcer volume fractions on thermal stress for Ti-6Al-4V/(TiC+TiB)

On the basis of Timoshenko's model[47], the failure mode of oxide scale is possibly the generation of micro-cracks at interface or the spallation of whole layer, which leads to the exposure of inner oxidized part into the high-temperature atmosphere. Owing to the usual process of oxidation of metals and alloys, the newly exposed region has high affinity with oxygen, which consequently favors successive chemical reactions. To provide a semi-quantified understanding on the effects of thermal stress induced spallation or cracks, a block-slot diffusion model addressed by Doilnitsyna[48] is applied to calculate the oxygen concentration distribution along the cross section. The mass-gain of the specimen during high-temperature oxidation is assumed to be contributed soley by the inward dissolution of oxygen, while no chemical reaction takes place during such a process. That is to say, the discrepancy between calculated and measured concentrations most probably represents the contribution of chemical reactions. The unidimensional oxygen concentration field corresponding to Neumann's boundary condition is expressed in the form of Gaussian error function[48]:

$$c(x,t) = c_0 \left[ erfc\left(\frac{x}{2\sqrt{Dt}}\right) - \exp\left(\frac{hx + h^2 t}{D}\right) erfc\left(\frac{x}{2\sqrt{Dt}} + h\sqrt{\frac{t}{D}}\right) \right] \quad (13)$$

where $x$, $h$, $t$, $D$, $c_0$ are diffusion distance, oxygen surface exchange coefficient, time, oxygen diffusion coefficient and oxygen concentration at the outermost location (set as 43.5 %, which is the average value of oxygen in $Al_2O_3$ and rutile-$TiO_2$). Because the oxygen diffusion coefficient for network-structured Ti-6Al-4V/(TiC+TiB) composites are currently in lack and the reinforcements content within the composites is relatively low, it is feasible as a first-order approximation to carry out such calculations using the oxygen diffusion coefficient of Ti-6Al-4V alloy measured by Guleryuz et al[35]. The physical parameters used in the calculations are listed in Tab. 5.

Fig. 14 illustrates the experimentally measured and calculated oxygen concentration distributions for Ti-6Al-4V/(TiC+TiB) composites oxidized at 873, 973 and 1073 K for 100 h. For the lower 873 K temperature, it is clearly seen in Fig. 14(a) that the measured oxygen concentration (in scatters with error bars) almost overlaps with that of theoretical prediction. The shadowed area in the figure shows the increment in oxygen concentration which might attribute to the following two reasons. At first, micro-cracks may appear in the composites but do not lead to marco-scale spallation. Secondly, oxygen could have stronger bulk diffusability within composites than the monolithic Ti-6Al-4V alloy, which makes the calculated results smaller than the actual measurements. When temperature rises up to 973 K, even though the variation tendency of experimentally measured oxygen concentration is similar to the results obtained by calculations, their difference becomes much larger than before. At 1073 K, a conspicuous discrepancy becomes apparent between the measured values and the calculated results. Considering all the mechanisms addressed before, this is caused mainly by the consequence of thermal stress induced mechanical spallation of oxide scales.

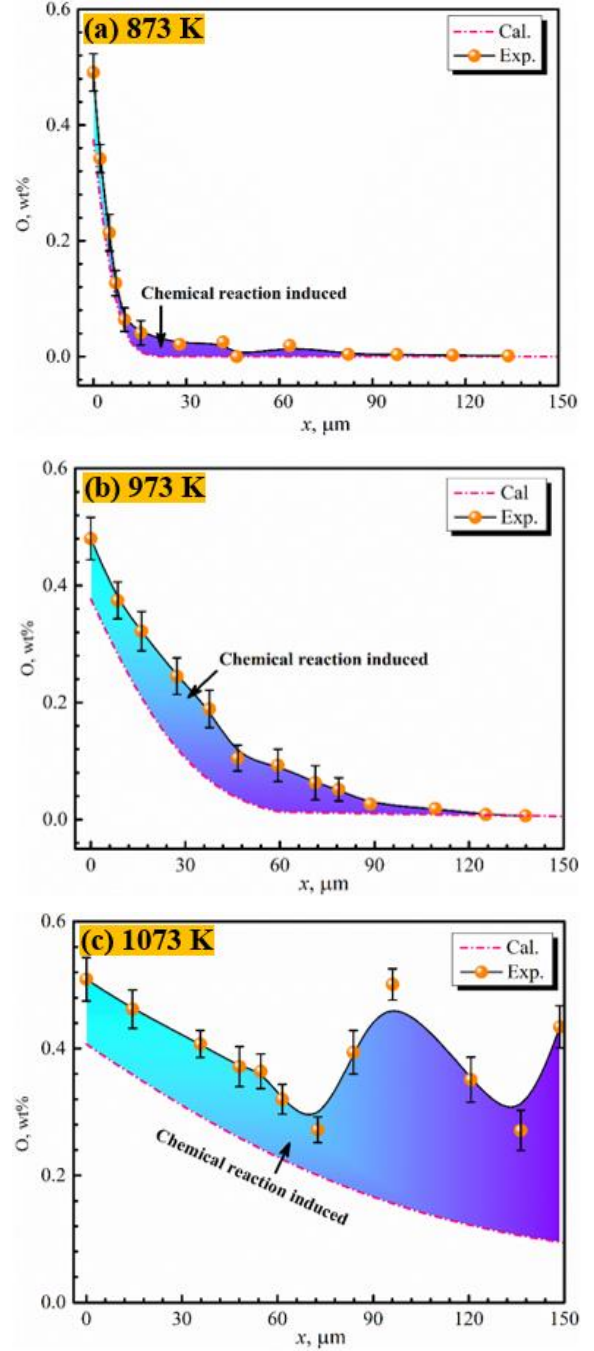

Fig. 14 Calculated and measured results of unidimensional oxygen distribution for Ti-6Al-4V/(TiC+TiB) oxidized for 100 h at different temperatures:
(a) 873 K, (b) 973K and (c) 1073 K

Tab. 5 Physical parameters used for predicting oxygen diffusion[35]

| Temperature | $D$, m²/s | $h$, m²/s |
|---|---|---|
| 873 K | $6.80 \times 10^{-17}$ | $10^{-8}$ |
| 973 K | $1.22 \times 10^{-15}$ | $10^{-6}$ |
| 1073 K | $1.28 \times 10^{-14}$ | $10^{-5}$ |

In the light of all the findings discussed above, a phenomenological model may be conceived to explain the growth of oxide scale over Ti-6Al-4V/(TiC+TiB) composites during cyclic oxidation. As depicted in Fig. 6(a), Al exhibits much more prominent oxidation tendency than Ti and V according to classical reaction thermodynamics. Simultaneously, the mobility of Al is two orders of magnitude higher than that of Ti at elevated temperatures. Both the thermodynamic predictions and the experimental results confirm that $Al_2O_3$ forms at the outermost part of the composites, leaving an Al depletion region in the sub-layer. Both Al and Ti near the surface will be completely oxidized to $Al_2O_3$ and rutile-$TiO_2$. Since the thermal stress is relatively low in this case, both types of oxide scale could well suppress the inner part from being successively oxidized, the corresponding mass-gain kinetics follows parabolic rule. When the total inherent stress (most of which is the thermal stress as discussed before) becomes large enough to stimulate the spallation or cracks, the unoxidized parts will be therefore exposed to high-temperature atmosphere, leading to extra chemical reactions, which produces the dual-layered oxide scale. If this process occurs repeatedly in every cyclic period, a multi-layered oxide scale will possibly form, as shown in Fig. 12 (c2). The mass-gain kinetic features will subsequently transfers into inconspicuous parabolic or linear trend, which are presented in Fig. s 7 and 8.

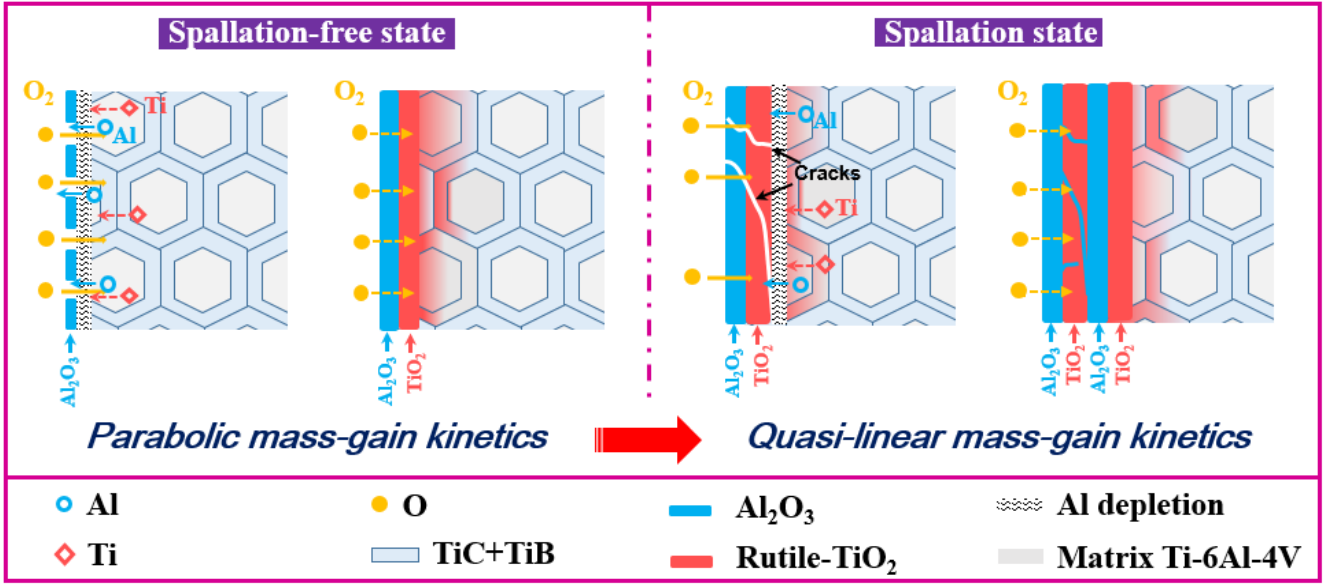

Fig. 15 Phenomenological model of oxide scale growth for Ti-6Al-4V/(TiC+TiB) composites from cross-sectional review

## 4. Conclusions

In summary, three dimensional network reinforced Ti-6Al-4V/(TiC+TiB) composites were successfully fabricated via reaction hot pressing techniques. Their cyclic oxidation resistance and concurrent structural evolution were experimentally investigated at 873, 973 and 1073 K for 100 h and also theoretically analyzed in details. The following conclusions are drawn accordingly.

(1) The introduction of network distributed in-situ TiC particles and TiB whiskers results in the refinement of α-phase in matrix Ti-6Al-4V alloy. The orientation relationship between TiB and α-Ti phase is extremely close to $(201)_{TiB}//(\bar{1}100)_{\alpha\text{-Ti}}$; $[11\bar{2}]//[0001]_{\alpha\text{-Ti}}$, while TiC and α-Ti phase show $(\bar{2}00)_{TiC}//(\bar{2}110)_{\alpha\text{-Ti}}$; $[001]_{TiC}//[01\bar{1}0]_{\alpha\text{-Ti}}$ orientation coincidence.

(2) The mass-gain kinetics characteristics reveal that the cyclic oxidation resistance of Ti-6Al-4V/(TiC+TiB) composites is superior to those of the monolithic Ti-6Al-4V alloy, Ti-6Al-4V/TiB and Ti-6Al-4V/TiC composites at all experimental temperatures. The reinforcements volume fractions are of greater significance to improve the oxidation resistance as compared to alloy matrix size.

(3) The mechanisms for the enhanced cyclic oxidation resistance of Ti-6Al-4V/(TiC+TiB) composites are possibly ascribed to the following aspects: (i) the introduction of TiC and TiB significantly decreases the thermal stress; (ii) the network-distributed reinforcements segment the oxide scale into smaller units; (iii) the well-assembled TiC particles become oxygen diffusion barrier; (iv) the TiB whiskers insert into oxide scale to provide extra fastening stress.

(4) Strong correlations exist between the variation of mass gain kinetics and the spallation of oxide scale or micro-cracks during cyclic oxidation. The combination of experimental results and calculations demonstrates that such effects become more obvious as temperature rises. A possible phenomenological model on the growth of oxide scale over Ti-6Al-4V/(TiC+TiB) composites is also proposed with the aid of classical thermodynamics and diffusion kinetics.

**Acknowledgements**

This work was financially supported by National Key R&D Program of China (No. 2017YFB0703100), National Natural Science Foundation of China under the grant numbers of Nos. 51671068 and 51471063. The authors are grateful to Assoc. Prof. G.H. Fan and Dr. X.P. Cui for their enthusiastic help with the experiments. S.L. Wei would also like to thank Assoc. Prof. J. Chang, Prof. W.L. Wang, Prof. W. Zhai and Assoc. Prof. Y.X. Xu for their stimulating discussion on cyclic oxidation kinetics.